# Requirements Engineering for General Recommender Systems

**Ivens Portugal**
David R. Cheriton School
of Computer Science
University of Waterloo
Waterloo, ON, Canada
iportugal@uwaterloo.ca

**Paulo Alencar**
David R. Cheriton School
of Computer Science
University of Waterloo
Waterloo, ON, Canada
palencar@cs.uwaterloo.ca

**Donald Cowan**
David R. Cheriton School
of Computer Science
University of Waterloo
Waterloo, ON, Canada
dcowan@csg.uwaterloo.ca

## Abstract

In requirements engineering for recommender systems, software engineers must identify the data that drives the recommendations. This is a labor-intensive task, which is error-prone and expensive. One possible solution to this problem is the adoption of automatic recommender system development approach based on a general recommender framework. One step towards the creation of such a framework is to determine the type of data used in recommender systems. In this paper, a systematic review has been conducted to identify the type of user and recommendation data items needed by a general recommender system. A user and item model is proposed, and some considerations about algorithm specific parameters are explained. A further goal is to study the impact of the fields of big data and Internet of things on the development of recommender systems.

## 1 Introduction

Recommender systems (RS) help users to choose or discover new content by monitoring and aggregating user's data, and displaying recommendations [72]. This research area has its origins in the mid-1990s [34] with the introduction of Tapestry [27], the first recommender system. With the evolution of technology and the popularization of the Internet, academia and industry adopted recommender systems for many application domains including, but not limited to, movies [62], books [61], tourism [48], social networks [67], academia (article recommendation) [87], and also technical domains such as programming (code segment recommendation) [25], and medicine (drug recommendations) [35].

Although the RS research field is around 25 years old and relatively mature, RS development is still a labor-intensive process. Software engineers must implement most parts of the system themselves with little or no automation. This is partly a consequence of the heterogeneity of information about a user or a recommendation item that the system must access. For example, to recommend a book, one recommendation approach may have access to previous book purchases of a particular user, whereas another approach may check which books were bought by other users that bought the first book. As a result, RS development becomes highly programmer-dependent, and hence, error-prone and expensive. However, with the advent of Big Data [91], RS development has become even slower.

One possible solution to these problems is the automation of recommender system development by defining a general recommender framework that can cover most, if not all, of the data that is required to recommend items for users. Such a framework has direct impact on requirements gathering, as software engineers have a pre-definition of the type of data required, and can be less focused on the RS architecture. Moreover, RS development becomes faster, less programmer-dependent, and cheaper, as it will be based on a general framework. The first step towards the creation of the general framework is the definition of a domain-independent user and item model capable of encompassing all types of information for a user and item.

This paper attempts to define a generic user and item model to be used in a framework for recommendation. A systematic review of the literature was executed to study what user and item information RSs use in the recommendation process. This paper also presents and describes a user and general item model, and concludes with a discussion of the results.

This paper is organized as follows: section 2 explains the main concepts about recommender systems. Section 3, describes the systematic review setting and section 4 discusses the results. This paper concludes in section 5 with a description of future work.

## 2  Recommender Systems

According to [72], the field of RS had its origins in cognitive science, approximation theory, information retrieval, forecasting theory, management science, and marketing. An RS can be defined as a system that can gather data from users, aggregate and process the data, and recommend an item. The first recommender system named Tapestry [27] appeared in 1992, and its developers used the term collaborative filtering to refer to their approach. The term, still in use today, is one of the three main categories of RSs namely: collaborative, content-based, and hybrid filtering [38].

Collaborative filtering refers to an RS property whereby the RS provides recommendations based on other users who have characteristics similar to the user that receives recommendations. As an example, consider the area of music. An RS with a collaborative approach may access a user profile containing relevant information such as age, gender, country, and preferred music genres, as well as the songs that the user listened to in the past, and compare this information with other users' profiles. The system can also create a user profile based on the ratings they gave to items (in this case, songs) and, in the same way, compare a user profile with other profiles. After calculating a similarity degree between users, the RS can discover other songs listened to by similar users, which might be of interest to the first user. However, RSs do not need to base their recommendation algorithm on the user data, but instead they can base it on the item data. Here, item data means information about the object of recommendation, such as movies, books, apps, drugs, and products or services. This approach is called content-based because recommendations are performed by gathering data about the item (or the content). An RS with a content-based approach may have access to data related to the item such as name, price, duration, or description, and may use it to calculate a degree of similarity between items. Once a user becomes interested in an item (e.g. a movie), the system displays similar items for that user.

One last approach, the hybrid, is the mixture of the two previous approaches. RSs with a hybrid approach benefit from user and item information to provide recommendations. For example, if a user is looking for apps on an app store and they become interested in an particular app, the RS can take into consideration the user's demographic information, age, gender, and previous purchases, as well as the app category, price, and description to provide the user with apps that similar users downloaded and apps that are similar to the one of interest to the user.

The process of gathering information from the user can be classified in two ways: explicit and implicit. Explicit user information gathering happens when the user provides data, such as when completing a form or answering questions. This approach is usually used to gather basic information about the user, such as their name, age, email, country, and user ratings. Conversely, implicit user information gathering happens when the RS monitors user behavior and obtains data from traces of information. As an example, consider that a user is searching for hotels on the Internet. The RS may have access to the browser, operating system and the country of the user because of the messages exchanged between the user's machine and the hotel company server. This approach raises several privacy issues, but the discussion about those is beyond the scope of this paper.

Besides the traditional recommendation process, some other forms of recommending can be found in the literature. Trust-based recommendations [50] rely on the information about the relationship between users. The concept is that recommendations made by family or friends have a higher impact than those made by unknown users. Trust between users is usually measured based on their relationship in social networks, which makes this domain more suitable for this recommendation approach. Context-aware recommendations [3] gather contextual information about the user and provide recommendations that differ depending on the current context. For example, a mobile app for music recommendation may have access to user's location and a device accelerometer to measure if the user is relaxing at home or working out at the gym. Depending on the context, the app may recommend a slow and calm, or fast-paced energetic song.

## 3  Systematic Review

One way to make RS development more independent of software developers, and consequentially reduce cost and development time is to create a framework for automatic RS generation. The first step in producing this framework is to represent information about users and items in a domain-independent model. These models must be general enough to capture the data that is used in real applications. Therefore, the goal of this systematic review is to find the set of information, related to users or to recommendation items, that is most used by real RS implementations described in the literature. This search is performed by answering the following research question (RQ):

- RQ: What information is used to model users and recommendation items in modern implemented recommender systems?

The decision to inspect only RSs that have been implemented is explained by the number of publications that introduce variations of recommendation algorithms, but either do not explain the user or item data gathered, or are never studied with real data. A second restriction of the systematic review is to consider only modern implementations of RS as software engineering tools and practices evolve. Inspecting recent implementations of RSs means analyzing the latest ways researchers are using user and item data. In summary, the study has two restrictions (R):

- R1: The study investigates RS approaches that have been implemented and tested with real or simulated data.
- R2: The study investigates modern RS approaches in the period from 2013 to 2015.

This systematic review applies criteria to filter results and improve its quality. Although it is difficult to predict what types of publications will be returned by a search query, the authors decided to define some exclusion criteria (EC) in advance. These criteria are:

- EC1: A publication must have been peer-reviewed and published in a conference, journal, book chapter, etc. Unpublished works are excluded.
- EC2: A publication must describe an RS approach sufficiently well that it is possible to identify user or item data the system needs. Publications that do not describe user or item data are excluded.
- EC3: A publication must describe a case study with real or simulated data. Publications that do not contain a case study description are excluded.
- EC4: A publication must be primarily in English. Publications in other languages are excluded.
- EC5: Books, letters, notes, and patents are excluded.
- EC6: A publication must be unique. If a publication is repeated, other copies of that publication are excluded.

Alternate terms for recommender such as the term "recommendation" are also considered. Moreover, the term "system" may be replaced by "platform" or "engine", and this is documented in the systematic review plan.

This systematic review has the following search query:

- TITLE-ABS-KEY((recommender OR recommendation) PRE/0 (system OR platform OR engine) AND ("case study")) AND (LIMIT-TO(PUBYEAR,2015) OR LIMIT-TO(PUBYEAR,2014) OR LIMIT-TO(PUBYEAR,2013))

This search query checks publication title, abstract and keywords looking for specific terms. The heterogeneity of terms to refer to RSs is expressed in the first part of the search query. Note that "PRE/0" is a search engine specific language denoting that the terms before it ("recommender" or "recommendation") must appear immediately before the terms after it ("system" or "platform" or "engine"). This search query also reflects the two restrictions of the systematic review. The first restriction (inspecting RSs that have been implemented) is expressed by the presence of the term "case study" in the publication title, abstract, or keywords. Searching for the term "case study" is one way of limiting the results to publications that have implemented and tested the algorithm they propose. The second restriction (inspecting modern RSs) is present in the search query by limiting the results to publications of the last three years: 2013, 2014, and 2015. Older publications may be considered in future studies.

The search query was used on the popular academic search engine Scopus [76] on July 7th 2015. It returned 93 publication entries that were then reviewed for quality. Some publication entries were discarded based on the exclusion criteria previously explained, and they are summarized on Table 3.1.

Table 3.1 - Publications that were included and excluded from the systematic review.

| Total retrieved | | 93 |
|---|---|---|
| Reason | Publications | Total |
| Conference / Proceedings entries | | 11 |
| Not about RS | [23] [29] [39] [53] [71] [78] [81] [95] | 8 |
| Not describing a case study | [57] [90] | 2 |
| Not sufficiently describing the approach | [13] [16] [66] | 3 |
| Not able to access | [7] [31] [41] [82] | 4 |
| Other language | [93] | 1 |
| Unreadable | [64] | 1 |
| Duplicated | [69] | 1 |
| Book | [50] | 1 |
| Publications retained | [1] [4] [5] [8] [9] [10] [11] [14] [15] [17] [18] [19] [20] [21] [22] [24] [25] [26] [28] [30] [32] [35] [37] [40] [42] [43] [44] [45] [46] [47] [48] [49] [51] [52] [54] [56] [58] [59] [60] [61] [62] [63] [65] [67] [68] [70] [73] [74] [75] [77] [79] [80] [83] [84] [85] [86] [87] [88] [90] [92] [94] | 61 |

As described in Table 3.1, 11 entries were conference proceedings and were excluded according to EC1. Out of the remaining publications, four were excluded because the authors could not access their content, even after asking colleagues, libraries, and using other search engines. The resulting 78 publications were downloaded. The authors excluded three more publications according to EC4 and EC6. One of these publications was in Chinese, a language that the researchers do not know. Another publication was unreadable, with badly formatted characters, and the last of these publications is the same version of [69] and was considered a duplicate. The researchers then read the abstract of all the 75 remaining publications, as well the approach and case study descriptions. The researchers then excluded 14 papers (=8+2+3+1) based on EC2, EC3, and EC5. In the end, the systematic review was performed with a set of 61 publications. The results of the review are described in the next section.

# 4 Systematic Review Results

While reading the publications for the development of this systematic review, the researchers decided to list some information about the RS approaches that could help explain the results, draw conclusions, or help in identifying new research opportunities. The first piece of information relates to data filtering approach. Recall that there are three categories: collaborative, content-based, and hybrid filtering. Table 4.1 below contains a summary of the results.

Table 4.1 - Numbers of publications that describe a particular data filtering approach.

| Approach | Publications | Total |
| --- | --- | --- |
| Collaborative filtering | [11] [17] [69] [73] [74] [75] [85] [88] [94] | 9 publications |
| Content-based filtering | [1] [5] [8] [10] [14] [15] [19] [20] [21] [24] [25] [26] [30] [32] [35] [37] [40] [42] [43] [45] [46] [49] [54] [56] [60] [61] [63] [65] [79] [80] [83] [84] [86] | 33 publications |
| Hybrid filtering | [4] [9] [18] [22] [28] [44] [47] [48] [51] [52] [58] [59] [62] [68] [70] [77] [87] [90] [92] | 19 publications |

One may note that the number of content-based RS approaches is greater than the sum of the other two, meaning that there might be a trend into looking for item data for recommendations. Similarly, the number of RSs using collaborative filtering is low, when compared to other categories. This means that most RSs approaches are not considering useful user profile data or user historical information when recommending. Although one possible explanation may be a privacy concern over user data, the number of RSs with a hybrid approach seems to refute this conclusion.

The authors also identified the domains in which RSs approaches were tested. It does not mean that the RS described in the publication was meant solely for this domain. Instead, the objective was to study what the most targeted domains are when validating RS proposals. The results are shown in Table 4.2 below:

Table 4.2 - Domains that RS approaches were tested on as well as the number of projects.

| Domain | Publications | Number of projects |
| --- | --- | --- |
| Movies | [17] [18] [28] [32] [37] [42] [51] [52] [54] [62] [63] [65] [68] [77] [79] [85] [92] | 17 |
| Technical (Software Engineering) | [8] [14] [24] [25] [25] [25] [25] [25] [25] [25] [28] [30] [40] [56] [73] | 15 |

| Academic / Professional | [17] [44] [49] [59] [74] [75] [77] [86] [87] | 9 |
|---|---|---|
| Tourism | [4] [9] [10] [48] [60] | 5 |
| Music | [17] [26] [58] [79] | 4 |
| Health / Medical | [1] [5] [35] [83] | 4 |
| Social | [45] [80] [90] | 3 |
| Hotel | [21] [22] [43] | 3 |
| Games | [25] [70] | 2 |
| Books | [61] [11] | 2 |
| Clothing | [19] [69] | 2 |
| Image processing | [25] [47] | 2 |
| Transportation | [46] | 1 |
| TV | [94] | 1 |
| Emotion | [84] | 1 |
| Humor / Jokes | [20] | 1 |
| Real estate | [88] | 1 |
| Taxes | [15] | 1 |

The previous table shows domain names followed by the number of projects that were validated in that domain. Note that the table does not show the number of publications about a particular domain, but projects. The researchers observed that some publications had either two or more descriptions of RSs approaches or they validated the same approach in two or more different domains. In both cases, the number of domains is not singular, and it is reflected in the domain count.

Based on Table 4.2, one may note that the movie, technical, and academic/professional domains have been used more often. The ease in accessing good data in these domains may explain their position in this ranking. The movie domain has two popular online movie databases that researchers can use: MovieLens [55] and IMDb [36]. Both databases contain data from users, movies and movie ratings. The technical domain refers to software engineering items, such as code, diagrams or artifacts in general. As RS is a field of computer science, researchers have easy access to these artifacts for validation of their approaches, and this can explain its second position in the ranking. The third most used domain, academic/professional, concerns items such as articles, white papers, or resumes. Again, researchers have access to a large number of scientific documents when doing research and this can explain the position in the ranking. When reading the 61 publications, the researchers were mainly concerned with identifying the user and item information that RSs used for their recommendations. For each publication, the researchers read the publication introduction and case study description. If the user/item information was not well described in those sections, the researchers then read the description of the approach, and finally any other relevant section. When user or item data was identified, the researchers noted this fact on a separate list. After the data gathering process, the researchers then moved to data analysis. In this step, the researchers tried to group user and item data into categories based on their apparent similarity. For example, user name, age and gender belong to a user profile classification, whereas user educational degree, place of education, and grade point average (GPA) belong to user education classification. Table 4.3 below depicts the user data categories that were chosen.

Table 4.3 - User data separated into categories.

| User category | Data |
|---|---|
| User general profile | Id, name, email, mobile phone number, birth date / age, gender, country, religion<br><br>Address, zip code, user location / position (device GPS), user previous path, user department, user laboratory<br><br>Marital status, family size, household income per month |
| User educational | Education / degrees, place of education, overall GPA |
| User professional | Occupation, industry, experience, seniority level, employed at, desired position, expected salary |
| User medical | User medication, user diagnostic |
| User social | User self-assessment / social network description, user trust statement (Facebook friends, Twitter following/followers, LinkedIn company following, LinkedIn connections), social network background color (Twitter, Facebook), link to social network, photo |
| User personality | User personality traits (assertiveness, cooperativeness, competing, accommodating, avoiding, collaborating, compromising) |
| User ratings | User ratings of items (number, start, like/dislike, comments/reviews), item currently browsing, items in shopping cart<br><br>User request note (similar to a feedback), list of published papers, user interest topics (based on published papers) |
| User system | Visit count, interaction with (links, buttons, maps, pictures), click log<br><br>User activity (accelerometer, inferences about whether user is idle, walking or running), environment (audio sample, inferences about whether user is on a meeting, restaurant, or at the street), time (inferences about morning, afternoon, evening)<br><br>Historical queries, video of the user, position of the eyes |

The great majority of publications use an identifier for the user such as a user id, name or email. However, there is a divergence when gathering user age. Some projects ask for the user data itself, while others require the user to complete a form with their birth date. Very little user data is historical, which supports the rationale behind the low number of collaborative approaches explained at the beginning of this section. There are few publications reporting case studies in the medical domain, which means that medical data about the user is not being gathered for recommendation. This is expected to change with the rise of wearable technology, and hence more case studies may be conducted in the near future. User social information is more concerned with trust-based recommendations as they take into consideration the relationship with users, such as friend connections.

Some publications attempt to categorize the user into a personality group, based on a questionnaire assessment, and then provide recommendations depending on the group to which they belong. The user ratings category concerns information about users expressing their preference for items, which can be done explicitly (number of stars or Facebook likes) or implicitly (items in shopping cart, user interest topic based on a historical metric). User system category relates to user data that can be gathered based on the system they are using. The information in this category is mostly implicit, because the system is sharing it with the RS.

To visualize data categories and their relationship with the user better, the researchers created the diagram shown in Figure 4.1. Only categories are shown as including data would only clutter the diagram and make visualization difficult.

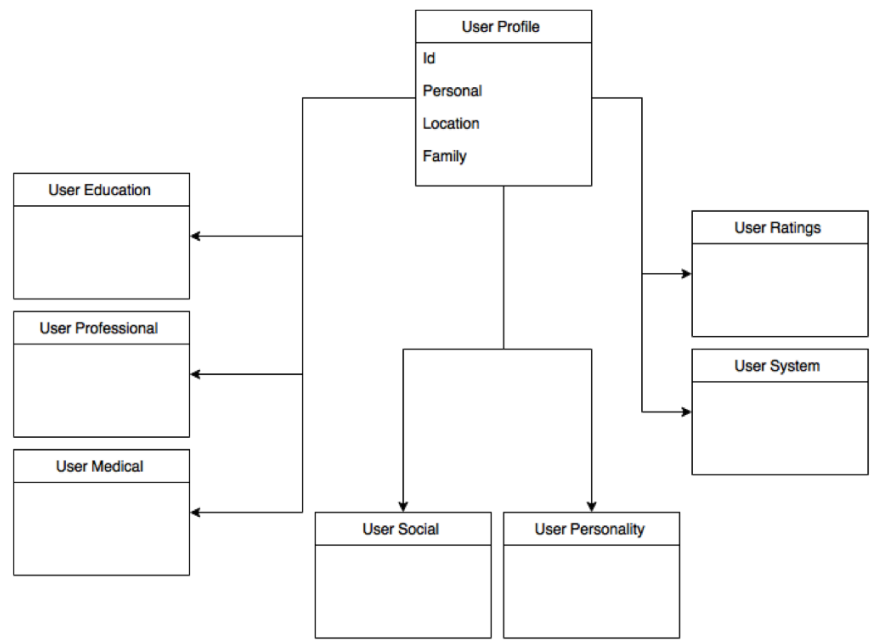

Figure 4.1. - A diagram showing the relationship among user data categories

While listing the user data identified from the 61 publications included in this systematic review, researchers also listed data about the items that RSs can recommend to users. Again, after all valid publications were read, the list with item data was analyzed and information was placed in categories based on the author's perspective. Table 4.4 and Table 4.5 show the results:

Table 4.4 - Item data related to identification.

| Item identification |
|---|
| Id, name, description, owner/author, year, keywords, URL |

Table 4.5 - Item data separated by types of information.

| Item attributes |
| --- |
| Number: citation, # of users that rated the item, rating, dose |
| Cost |
| Duration |
| Text: words in document, code |
| Location: country, region, city, city area |
| Distance from item to user |
| Boolean: isOpen, isIndoor, isAbleToVisit |
| Image |
| Scores: cleaning, location, service level |
| Classification / type: (residence, hotel, …) (music genre) |
| List of sub items:<br>• References, class and method names, namespace, number / location of XML tags<br>Actors, directors, producers, courses offered |
| Usage: number of times used, relative usage, relative importance |
| User x Item relationship: companies you may want to follow |

Item data was divided into two main classifications: identification and attributes. The first classification regards data for identification such as id, name, and URL. Again the majority of the publications described a form of identifying an item using its id or name. The second table depicts the possible attributes of an item. Some data was grouped based on what type of information they represent. Items can have numbers associated with them, such as number of citations, or number of users that rated that item. Cost and duration are also numbers. However, as they are widely used in several domains of RS, they were listed separately. Items can also have associated text as the words in a document, or program code. Item location and distance from the user may refer to a country, region or city, and how far it is from the user location, especially if the item is a point of interest (POI). Boolean information for items refers to yes/no attributes such as whether the item is currently open, or if it is possible to visit it. Some publications used scores given by the user, which are different from user ratings, for some characteristics of the item, such as its location or service level. Items can have a type or a classification, as movies and songs have a genre. Items can also have a list of sub items related to them. For example, academic papers have a list of references, and a movie has a list of actors. The usage category on an item attribute describes information about a manipulation of the item. Some publications used the number of times that an item was used, while others calculated a relative usage (in relation to all other items) or importance. Lastly, one item attribute can be the relationship between the user and the item. The RS can use this relationship to recommend items to other users. For example, LinkedIn recommends that users might wish to connect based on their mutual interest in a company.

Similar to user data, item data was modeled in a diagram for better visualization. Only item categories were written again to avoid clutter. The result is shown in Figure 4.2.

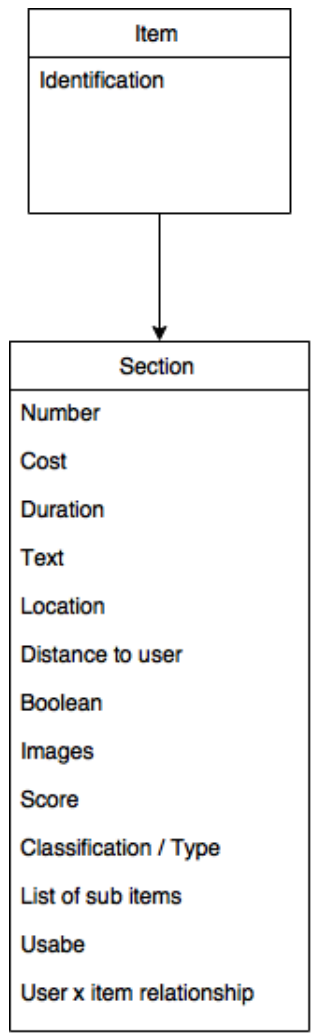

**Figure 4.2 - A diagram showing the realtionship between item data and its categories.**

Both user and item models were systematically gathered from RS approaches that have been implemented and tested in a case study. These models can certainly be expanded, but they already represent a good starting point in requirements engineering for recommender systems. The researchers expect to have answered the research question with the development of these models.

When reading publications for a systematic review, the researchers often draw other conclusions or observe some trends that are not directly related to the research question. This is worth documenting, so that a study may further enrich the knowledge in the area. In the next paragraphs, some conclusions that are indirectly related to the main purpose of this systematic review are described as well as some trends.

Any RS needs to have access to either user or item data (or both) on which to base recommendations. However, in some cases, only raw data is available. For example, the system may need the user age, but only have access to the birth date. Therefore raw data (birth date, in this example) must be processed before being used by an RS, denoting the two layers of information depicted in Figure 4.3. More complex examples are counting the number of occurrences of a particular word in a document or obtaining the predominant color of a photo.

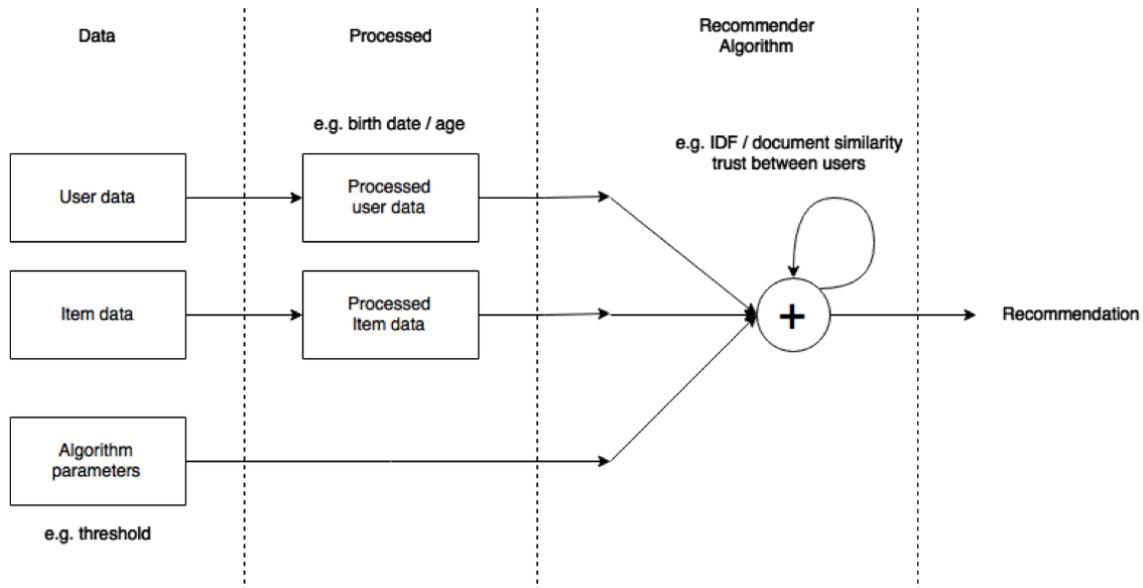

**Figure 4.3 - Two layers of information meaning that user or item data may be processed before being used by recommender systems.**

The authors also observed that RSs need some parameters that are algorithm-specific to perform recommendations well, such as a threshold, the number of nodes or the depth of a tree. This information, also described in Figure 4.3, is not modeled with user or item data, but is necessary when developing RSs. Finally, some RS approaches generate metrics that are used to improve recommendation. For example, when recommending products on a social network, an RS may use a trust-based approach and generate a trust level between users to recommend only products in which trusted friends are interested. This is also modeled in Figure 4.3, under "recommender algorithm".

Although mapping all data that RSs require is difficult, research in this direction is necessary because of the emergence of the field of big data, in which huge amounts of data can be accessed from different sources and in different formats. Software engineering is improving to embrace this new reality. The discussion presented here can be extended to the big data field, but more studies should be done to evaluate the impacts of big data.

# 5  Conclusion & Future Work

The goal of recommender systems (RS) is to provide the user with items or data of possible interest that will help the user make a more informed decision. Recommendations are made after gathering (explicitly or implicitly), processing, and analyzing user or item data. However, developing an RS is a manual and labor-intensive task for programmers, because of the apparent heterogeneity of data available to be processed. One way to facilitate requirements gathering, and consequently RS development, is the definition of a general user and item model that can adapt to any domain. This study describes a systematic review whose goal is to identify all user and item information used in modern and implemented RSs.

After reading the publications included in the systematic review, the authors concluded that there is a significant number of RS case studies (61) published in the last three years. The majority (33 publications) use content-based filtering, which means that the recommendation process is based on the item data, such as price. There is a low number of RSs with collaborative filtering, which may indicate that, either comparison of user profile or user historical information has not been investigated enough. Privacy concerns over user data may explain this result. However the number of hybrid approaches (19) shows that this concern is not preventing studies with user data.

There are several domains in which RSs can be used and more are yet to be discovered. However, the researchers observed that the three most used domains for recent publications are movies, technical, and academic/professional. The ease of access to real data appears to be an important factor for this result, since researchers have access to rich databases, such as MovieLens, IMDb, academic and conference search engines, and code.

In addition, this study resulted in the creation of a user and an item model to be used in the development of RSs. The user model, fully described in the last section, classifies user information into 8 categories: profile, education, professional, medical, social, personality, ratings, and system. The category names clarify what information they depict. The last two categories (ratings and system) concern user ratings of items and implicit information about the user behavior, such as a log of mouse clicks. Similarly, the item model separates information into two main categories: identification and attributes. The first contains information that identifies the item, such as its id or name. The second category is divided into several types of data, such as number, text, or lists that contain the data (quantity, words, and references).

The authors also identified that some user or item information may be processed before being delivered to the RS, such as distance between users derived from each user's location. Moreover, recommendation algorithms may require an implementation-specific parameter (e.g. threshold), and some approaches will calculate those parameters themselves, such as the TF-IDF algorithm [33].

The results presented in this work can be extended, including more information about new case studies with different technologies. One major shift in computer science that must be relevant in this context is the emergence of Big Data, which is composed of very large datasets, whose data has several formats, and is constantly changing. It may change how user and item data are modeled, and especially, how information is gathered from users and items in requirements engineering. A Scopus search (at the time of the development of this systematic review) of the terms "Recommender System", "Big Data", and "Case Study" returned only one publication, which shows that more work with practical results needs to be done. Another computer science field that is closely related to Big Data and can also play a significant impact in the development of RSs is the Internet of Things (IoT) [6]. It is the relationship between the user and a set of gadgets, such as smartphones, smartcars, smarthouses, or wearables. All devices can produce very large sets of information and studies in this direction are required to understand better the impact of this data in recommendations.

# 6  Acknowledgments

The authors would like to thank the Natural Sciences and Engineering Research Council of Canada (NSERC), the Ontario Research Fund of the Ontario Ministry of Research and Innovation, SAP, and the Centre for Community Mapping (COMAP) for their financial support to this research.